# Switchable Adhesion Actuator for Amphibious Climbing Soft Robot

Yichao Tang\*, Qiuting Zhang, Gaojian Lin, and Jie Yin\*

Applied Mechanics of Materials Laboratory, Department of Mechanical Engineering, Temple University, Philadelphia, PA 19122, USA

\*Email: yichaotang@temple.edu (Y. T) or jieyin@temple.edu (J. Y)

Abstract: Climbing soft robots are of tremendous interest in both science and engineering due to their potential applications in intelligent surveillance, inspection, maintenance, and detection under environments away from the ground. The challenge lies in the design of a fast, robust, switchable adhesion actuator to easily attach and detach the vertical surfaces. Here, we propose a new design of pneumatic-actuated bioinspired soft adhesion actuator working both on ground and under water. It is composed of extremely soft bilayer structures with an embedded spiral pneumatic channel resting on top of a base layer with a cavity. Rather than the traditional way of directly pumping air out of the cavity for suction in hard polymer-based adhesion actuator, we inflate air into the top spiral channel to deform into a stable 3D domed shape for achieving negative pressure in the cavity. The characterization of the maximum shear adhesion force of the proposed soft adhesion actuator shows strong and rapid reversible adhesion on multiple types of smooth and semi-smooth surfaces. Based on the switchable adhesion actuator, we design and fabricate a novel load-carrying amphibious climbing soft robot (ACSR) by combining with a soft bending actuator. We demonstrate that it can operate on a wide range of foreign horizontal and vertical surfaces including dry, wet, slippery, smooth, and semi-smooth ones on ground and also under water with certain load-carrying capability. We show that the vertical climbing speed can reach about 286 mm/min (1.6 body length/min) while carrying over 200g object (over 5 times the weight of ACSR itself) during climbing on ground and under water. This research could largely push the boundaries of soft robot capabilities and multifunctionality in window cleaning and underwater inspection under harsh environment.

**Introduction**

Recently, the study of soft robots has attracted tremendous research interest in both science and engineering, due to their great potential to interact with humans and the environment more safely and more adaptively. Soft continuum robots are often made of highly deformable soft materials to allow compliance, endurability, and elasticity[1]. Bioinspired by soft-bodied systems in nature such as caterpillar[2, 3], meshworm[4, 5], octopus[6-8], and fish[9], extensive research efforts have been dedicated to mimicking a variety of biological locomotion modes, including walking[1, 2, 4, 5, 10-12], jumping[3, 13, 14], and swimming[9, 15-17] in soft robotics. The locomotion is realized through deforming their soft bodies by means of bending[11, 18], expanding or contracting [4, 19], twisting [2, 5] or combined in response to external stimuli, including pneumatic or hydraulic pressure[1, 11], temperature[4], light[2], and electrical[3] or magnetic field[20].

Despite these advancement in locomotive soft robotics, design of amphibious soft climbing robots on ground and under water remains to be explored, the study of which could largely push the boundaries of robot capabilities and multifunctionality[21]. Climbing, as one of the most fundamental locomotion modes in nature, has long been fascinating to the researchers in the biological and robotics field due to its potential applications in intelligent surveillance, inspection, maintenance, and detection under environments away from the ground [22]. Given the harsh working environment (e.g. high altitude), when compared to rigid robots, climbing robots made of soft materials could largely increase the chance of surviving a fall due to their extreme compliance. To counter the gravity force, the main challenge of self-supported climbing robots lies in the design of fast, switchable, and strong adhesion actuators for not only easily attaching and detaching the targeted vertical surfaces upon actuation for locomotion, but allowing certain load-carrying capability for potential functionality under different complex 2D or 3D working environment[23]. In rigid climbing robots constructed from rigid links and adhesion pads, the adhesion to surfaces has been achieved through two major mechanisms: gecko inspired micro-fibrillar adhesives for dry and directional adhesion [21, 23-26] and octopus inspired suckers for vacuum suction adhesion[27-31]. However, it remains very challenging to either apply the gecko-inspired adhesives for wet adhesion and underwater locomotion, or directly transfer air-pumping suckers to design soft climbing robotics due to the potential issue of structure failure. For example, vacuum pumping a suction cup made of extremely soft materials (e.g. ecoflex, elastosil, hydrogel etc) can easily lead to the collapse of the soft structure[32, 33], thus fails to achieve desired strong and stable adhesion upon actuation.

To address the challenge, here, we propose a novel soft adhesion actuator that allows for switchable and strong dry/wet adhesion without structural collapse upon rapid actuation. The soft adhesion actuator is made of extremely soft elastomer ecoflex and composed of two layers with an embedded spiral pneumatic channel on top of a cylindrical chamber. Rather than the traditional way of applying negative pressure for suction adhesion (*i.e.* pumping air out of the chamber for vacuum), we use positive pressure to deform the planar bilayer structured soft actuator into an inflated 3D domed shape for achieving stable and switchable adhesion. Guided by a simplified bilayer doming model, we conduct a parametric study on manipulating the geometry of the adhesion actuator for achieving high and stable shear adhesion force. Based on this adhesion actuator, then we design and fabricate an inchworm-inspired amphibious soft robot that can climb and walk on ground and under water. We demonstrate and characterize its wide capability of vertical climbing on various types of surfaces with certain load-carrying capability, including smooth, semi-smooth, dry, wet, and slippery surfaces, as well as underwater walking and climbing.

**Results and Discussions**

*Working mechanism of soft adhesion actuators*

Figure 1a shows the schematics of octopus suction cup inspired design of a switchable adhesion actuator. It is composed of a bilayer structure with an embedded pneumatic spiral channel on the top and a cylindrical chamber (cavity) underneath (right of Fig. 1a). The soft actuator is fabricated by curing elastomer (Ecoflex 00-50, Smooth-on Inc.) in 3D printed molds followed by demolding (See Supporting Information for details). Fig. 1b shows the fabricated soft adhesion actuator before (left) and after actuation (right) by depressurizing/pressurizing the spiral pneumatic channel on the top, respectively.

The working mechanism for actuating the switchable adhesion is schematically illustrated in Fig. 1c. Octopus-sucker exhibits strong adhesion by generating pressure difference between the cavity inside the suckers and outer circumstance upon muscle contraction to squeeze out the air in the cavity. Rather than squeezing air or water out of the cavity in the octopus sucker, we inflate air into the channel on the top to generate a negative pressure in its underneath cavity for achieving adhesion. After inflating air into the spiral channel on the top, it will generate mismatched expanding deformation between the two layers and deform the planar structure into a 3D dome shape after actuation[34, 35] (right of Fig. 1b and Fig. 1c). Consequently, the connecting underneath chamber (cavity) deforms coherently into a dome shape with an increased volume of $(V_o + \Delta V)$ ($V_o$ is the original volume of the chamber or cavity and $\Delta V$ is the volume change). For the case of on ground, before actuation, the cavity is filled with atmosphere air and its pressure inside is equal to the atmosphere pressure $P_o$. Upon actuation, the increased volume leads to a pressure drop $\Delta P$ in the cavity in terms of $P_o V_o = (P_o - \Delta P)(V_o + \Delta V)$, thus, it generates a pressure difference $\Delta P$ between the internal cavity $P_c$ and the external circumstance $P_o$, i.e. $\Delta P = P_o - P_c$, forcing the soft actuator to conformably attach the foreign surface with its extreme compliance. The generated adhesion force is mainly determined by the pressure difference $\Delta P$ inside and outside the cavity, which can be tuned by manipulating the geometrical size of the actuator, including the channel size, the layer thickness above ($h_0$) and underneath ($h_1$) the spiral channel, and the cavity volume (Fig. 1c).

Based on this simple method of generating adhesion force via doming-induced pressure difference, we can rapidly and reversibly switch on and off the adhesion actuator by simply pressurizing and depressurizing the embedded spiral channel, respectively. It should be noted that the traditional suction actuator, by means of directly pumping air out of the cavity for vacuum-induced adhesion, will lead to the collapse of the soft structure structure[32, 33] due to its highly deformable and extreme compliance characteristics. In contrast, the positive pressure based soft adhesion actuator is more stable and controllable. Pneumatic inflation into the soft adhesion actuator can help stiffen the soft structure, thus retains and enhances the structural stability of the suction cup.

*Modeling of adhesion actuator as simplified bilayer doming system*

To shed some light on the design of the adhesion actuator, we employ a simplified and approximate bilayer model with nonuniform axisymmetric mismatched expansion between circular-shaped layers [36] to understand the volume change in the cavity after doming.

For a bilayer system composed of a circular thin film (thickness of $h_f$) on a substrate (thickness of $h_s$) with radius of R as shown in Fig. 2a ($h_f \ll h_s$), when it is subjected to a nonuniform but axisymmetric misfit strain $\varepsilon_m(r)$ along the radial direction $r$, the height $u_z$ of the deformed dome structure along the normal direction $z$-axis can be obtained as [37]

$$u_z = -\frac{6E_f h_f}{1-v_f^2} \frac{1-v_s^2}{E_s h_s^2}(1+v_s)\left[\int_0^r \frac{1}{r}\int_0^r \eta\varepsilon_m(\eta)d\eta dr + (\frac{1-v_s}{1+v_s} - 2\frac{v_s-v_f}{(1+v_s)^2})\frac{r^2}{2R^2}\int_0^R \eta\varepsilon_m(\eta)d\eta\right] + A \quad (1)$$

where $E$ and $v$ are the Young's modulus and Poisson's ratio. The subscripts "$f$" and "$s$" represent the film and substrate, respectively. $r$ (or $\eta$) is the radial distance from the center. $A$ is a constant to be determined by satisfying the assumed boundary condition of $u_z(R) = 0$. When applying the continuum bilayer model to the adhesion actuator as shown in Fig. 2b, we assume that $h_f$ and $h_s$ take the value of ($h_1 - h_0$) and ($2h_0 + h_c$), respectively, i.e. $h_f = h_1 - h_0$ and $h_s = 2h_0 + h_c$. We also assume the same materials properties, $E$ and $v$, for both the film and substrate. Here, we take a simplified homogenized bilayer model by neglecting the detailed channeled structure in the top layer while considering the mismatched expansion induced by pressurization in the channel. Then Equation (1) becomes:

$$u_z = -6(h_1-h_0)\frac{1+v}{(2h_0+h_c)^2}\left[\int_0^r \frac{1}{r}\int_0^r \eta\varepsilon_m(\eta)d\eta dr + \frac{1-v}{1+v}\frac{r^2}{2R^2}\int_0^R \eta\varepsilon_m(\eta)d\eta\right] + A \quad (2)$$

Upon expansion of the top layer, the volume change of the cavity can be obtained as:

$$\Delta V = 2\pi \cdot \int_0^R r u_z dr \quad (3)$$

It should be noted that Equation (3) ignores the displacement along the radial direction since this displacement is significantly small when the thickness of the cup (the thin wall that wraps around the cavity) is large, which will limit the radial expansion of the whole structure. In addition, Equation (3) is an idealized situation, which does not consider the effect of the resulting internal pressure drop inside the cavity from the volume change. A more refined bilayer bending model will be developed in the future by balancing the in-plane mismatched expansion induced doming deformation and the doming-induced pressure change in the cavity.

With Equation (3), then we can determine the pressure difference between the cavity of air and atmosphere upon actuation as below:

$$\Delta P = \frac{P_o \Delta V}{V_o + \Delta V} \tag{4}$$

Utilizing Equations (2) – (4), the pressure change of the cavity for a bilayer doming system upon actuation can be predicted. It can be seen that the expansion difference $\varepsilon_m(r)$ between the two layers plays a dominant role in determining the pressure change.

To determine the expansion $\varepsilon_m$ along the radial direction in our adhesion actuator, we use the digital image correlation (DIC) to track the expansion of the top layer and thus quantify $\varepsilon_m$ as a function of $r$ upon inflation. Fig. 2c show the DIC image on the radial strain contour of the top-view adhesion actuator, where the measured expansion coefficient as a function of the radial position upon 4mL air inflation is plotted in Fig. 2d. The measurement shows that $\varepsilon_m$ varies significantly along the radial direction despite the constant height of the channel, where the expansion rate increases nonlinearly from zero in the center to arrive its peak at $r/R \approx 0.8$, and then decreases to approximately zero at the edge of the actuator. After substituting the fitted experimental curve of $\varepsilon_m$ into Equation (2)-(3) (red dashed line expressed by $\varepsilon_m = -0.75\bar{r}^3 + 0.817\bar{r}^2 - 0.05\bar{r} + 0.005$ with $\bar{r} = r/R$), the theoretical volume change of the adhesion actuator upon actuation can be predicted as $\Delta V/V_o = 0.16$, which agrees very well with our measured volume change upon 4mL inflation: $\Delta V/V_o = 0.14 \pm 0.02$ despite the simplified model.

*Effect of the geometric parameters on adhesion*

Our pneumatic adhesion actuator itself is a complicated system and its adhesion behavior is determined by a couple of characterized geometric parameters. In this research, among them, we focus on three major parameters governing the doming deformation of an adhesion actuator with a given radius R and air channel size, namely, the layer thickness $h_1$ between the spiral channel and the cavity and the height of the cavity $h_2$ (Fig. 1c), which mainly determine the deformation governed volume change in the cavity, as well as the "density" of the spiral channel for a high expansion by manipulating the distance between the channel. The strength of the adhesion actuator is quantified by measuring the maximum shear adhesion force $S_{max}$ through pure shear testing as schematically illustrated in Fig. 3a. $S_{max}$ is defined as the critical pull-off shear force to detach or slide along the substrate surface. Fig. 3b shows the result of the measured $S_{max}$ of the soft adhesion actuator on a smooth acrylic surface as a function of the volume of input air into the spiral pneumatic channel. As the volume of inflated air increases, the soft adhesion actuator deforms gradually into a dome shape with an increasing dome height. The measured maximum shear adhesion first increases monotonically and then approach a plateau with an actuation pressure of 62kPa. In the following experimental test, the measurement of the shear adhesion force is conducted by attaching the actuator to the same smooth acrylic surface at the same actuation pressure of around 62kPa.

The results of the parametric studies (Fig. 3c-3f) show that to achieve both a high and robust adhesion force, a moderate value of $h_1/h_0$ and $h_2$, as well as a relatively higher number of revolutions in the spiral pneumatic channel is recommended. When $h_2 = 2.7$ mm is fixed, Fig. 3c shows that as $h_1/h_0$ ($h_1/h_0 \geq 1$) increases, $S_{max}$ increases first and arrives at its peak value at $h_1/h_0 = 1.35$, then it decreases with further increase of $h_1/h_0$. It is reasonable that when $h_1/h_0$ is relatively small and close to 1, upon pressurization, the deformation in the bilayer structure is dominated by the structural radial expansion rather than doming, leading to a small out-of-plane doming height $u_z$, as evidenced by Equation (2), thus a weak adhesion. However, when $h_1/h_0$ becomes relatively large, it will require more inflation to bend the thicker envelope of the cavity. Similarly, it will also result in a weak adhesion due to the relatively increased bending stiffness of the bottom cavity structure. Thus, in the following, we choose $h_1/h_0 = 1.35$ for our adhesion actuator due to its exhibited largest shear adhesion force.

Regarding the effect of the cavity volume, Fig. 3d shows that a smaller cavity volume (small $h_2$) leads to a larger shear adhesion force while keeping other parameters constant. It can be explained as below: we assume that the volume change ($\Delta V$) of the cavity is mainly determined by $h_0$ and $h_1$ rather than $h_2$. A relatively smaller initial cavity volume ($V_o$) produces a larger value of $\Delta V/V_o$, thus a larger pressure difference $\Delta P$ in terms of Equation (4) and a higher $S_{max}$. We find that the strongest adhesion is achieved by setting the actuator without a cavity on the bottom, i.e. $h_2 = 0$. Despite the strongest adhesion, a potential issue of unstable contact with the target surface exists for the case without a cavity. We observe that upon actuation, this non-cavity-based adhesion actuator has less contact with the substrate (Fig. 3e) when compared to those with a relatively larger cavity space, which makes it difficult to firmly conform to the substrate surface, especially on semi-smooth surfaces, thus it becomes more susceptible to potential air leaking. Therefore, a balance between the good conformability of the adhesion actuator and its adhesion force upon pressurization should be considered for the design of climbing soft robots (here we choose $h_2 = 2.7$ mm for a good balance).

Similarly, to achieve a high adhesion force, we can increase the "density" of the spiral channel. Here the "density" of the spiral channel is defined as the ratio of the volume that the spiral channel occupies with respect to the volume of the adhesion actuator. Since this "density" can be determined by multiple geometric parameters, for simplicity, here we mainly vary the number of revolutions for the spiral channel within the same unit volume (indicated by yellow in the inset of Fig. 3f) by tuning the distance between the spiral while keeping other geometrical parameters unchanged. Fig. 3f shows that, upon the same actuation condition (62kPa), as the number of revolutions increases, the maximum shear adhesion increases monotonically and then approach a plateau. Thus, a relatively larger number of 4 revolutions is chosen for the spiral shape to ensure the good adhesion performance of our adhesion actuator in the following.

*Amphibious Climbing Soft Robot (ACSR)*

Equipped with the information of designing the adhesion actuator with high and robust adhesion forces, next, we use the adhesion actuator to design an amphibious climbing soft robot by combining with a soft bending actuator for locomotion on ground and under water.

Fig. 4a schematically shows our bio-inspired design of the ACSR by mimicking the locomotion of an inchworm[18]. The fabricated ACSR under an actuated and bended state is shown in Fig. 4b. Similar to an inchworm, ACSR consists of three actuation parts (Fig. 4a): two adhesion actuators on both ends mimicking the head and tail of an inchworm, which enables a strong switchable adhesion force required for attaching and detaching on target surfaces as discussed above; one classic pneumatic bending actuator [1, 38] in the middle to mimic the inchworm's bendable soft body for the locomotion purpose. It has embedded rectangular wave-like pneumatic channels in the hyperelastic elastomer (highlighted in blue in Fig. 4b) and bonded with a strain-limiting layer (highlighted in gold color in Fig. 4b, Staples® card stock paper) on its bottom. Combined with the adhesion actuator, the bending actuator will drive the locomotion of the soft machine via bending/unbending its soft body upon pressurization/depressurization.

In addition to the three actuators, a hard polylactide (PLA) plastic slider (Fig. 4b and Fig. S3) is built to connect the two adhesion actuators. The connector will allow the two adhesion actuators to translate and move freely along the slider within the same plane only. Meanwhile, it will constrain their possible rotation movement to prevent their detachment or fall from the target surface, especially for climbing vertical surfaces (Fig. S4). Thus, the slider will force the two adhesion actuators to firmly contact the substrate during the walking and climbing locomotion. Furthermore, the slider plays another role as a hard skeleton, providing the soft machine with enough support and stability.

By pneumatically actuating the three actuators in sequence with a pneumatic control system (Fig. S5), we demonstrate both the walking (Fig. 5a, Supporting Video S1) and climbing (Fig. 5b, Supporting Video S2) modes of the designed soft machine on a smooth and dry surface (e.g. an acrylic plate) with a certain load-carrying capability. One cycle of the locomotion of the ACSR involves five sequential steps. First, its "head" is pressurized (~62kPa) to adhere to the target surface (Fig. 5a(i) and Fig. 5b(i)). Second, its soft "body" is then activated and becomes bended by pressurization (~100kPa) to move its "tail" and pull the carried load forward (horizontal surface, Fig. 5a(ii)) or upward (vertical surface, Fig. 5b(ii)) for a load-carrying locomotion. Third, its "tail" is actuated through pressurization (~62kPa) to attach to the substrate, which will help to hold the carried load (Fig. 5a(iii) and Fig. 5b(iii)). Fourth, the "head" is switched to an adhesion-off state by depressurization to release the adhesion (Fig. 5a(iv) and Fig. 5b(iv)). Last, we depressurize and unbend the soft "body" to release the stored bending energy to push its adhesion-off-state "head" to move forward along the slider (Fig. 5a(v) and Fig. 5b(v)). Simply repeating the sequential steps above can achieve an effective locomotion with a large translation distance (After one cycle, the soft robot can move a distance of about 38 mm. The locomotion speed is about 286 mm/min). The motion of the soft machine can be actively tuned and controlled by varying the pneumatic pressurization and adjusting geometric parameters of the slider.

Compared to previously reported soft machines[39], one of the advantages of the proposed soft robot is that it can carry more loads with the help of the adhesion actuator. For example, on a horizontal flat surface (Fig. 5a), our robot can drag a 350g steel bar (we didn't use wheels to decrease the friction) or even heavier object forward (Supporting Video S1). When the soft robot climbs on a vertical surface (acrylics), it can easily lift up a 200g steel bar (Fig. 5b, Supporting Video S2).

*Climbing soft robots on multiple types of surfaces*

In addition to its climbing on smooth and dry surfaces (e.g. the acrylic sheet), we further examine its climbing capability on multiple different types of surfaces, including semi-rough surfaces (e.g. rough sandpaper and indoor painted wall) as well as wet or slippery lubricated surfaces. The demonstrated proof-of-concept videos in the supporting information show that the proposed climbing soft robot can even carry a load of 200g to climb on a variety of surfaces, including dry, wet, slippery, and semi-rough surfaces (Fig. 6a, Supporting Video S2 and Supporting Video S3).

The maximum load-carrying capacity of the soft climbing machine is mainly determined by the maximum shear adhesion force $S_{max}$ of the adhesion actuator. To better understand its load-carrying capacity and climbing capability, we conduct the measurements of the generated maximum shear adhesion force $S_{max}$ on various surfaces, including acrylics, glass, steel, and paper. All the measurements were conducted under the same actuation pressure of 62kPa with a corresponding pressure of ~0.476atm inside the cavity. The measurement results (Fig. 6b) show that different types of the substrate materials do not significantly affect the maximum adhesion force that the adhesion actuator can generate. The measured adhesion force ranges from about 6N to 8N on all the measured dry and smooth substrates, indicating a strong loading capacity of carrying objects that are 60-80 times the weight of adhesion actuator itself (10g) on vertical smooth surfaces.

For wet surfaces, experimental result shows that the shear adhesion force of the adhesion actuator on wet acrylics (~6.96N) is slightly smaller than that on dry acrylics (7.97N), thus, it can still function under more challenging circumstance, for example, outdoor performance during the rainy days. For slippery surfaces, such as acrylics surfaces sprayed with lubrication liquid (PVA Release Film, Fiber Glast Development Corporation), the measured maximum shear adhesion force shows a much larger error deviation, which is mainly attributed to the amount of lubrication liquid sprayed on the surface. For a tested acrylics surface (surface area = 64cm$^2$) sprayed uniformly with 0.4g lubrication liquid, $S_{max}$ is measured to be 6.23N, which is close to its performance on dry acrylics. The exhibited strong adhesion of the proposed adhesion actuator accounts for the soft robot's climbing on wet and even slippery surfaces, which remains very challenging for conventional locomotive soft robots without adhesion actuator due to their low surface friction.

Next, the possibility of the soft robot's climbing on semi-smooth surfaces is further examined. Here, semi-smooth surfaces are defined as lightly rough surfaces with the roughness amplitude (arithmetical mean height $S_a$) smaller than 20μm. As a proof of concept, we use a lightly rough sandpaper (Norton, grit number =180, $S_a$ =17.43μm) as the targeted climbing semi-smooth surface. The maximum shear adhesion force of the adhesion actuator on the sandpaper is measured to be 8.95N, which is even larger than those on smooth surfaces after the bottom surface treatment of the adhesion actuator. This enhanced adhesion is due to the relatively larger friction of semi-smooth surfaces compared to that of smooth surfaces. With this strong adhesion, we successfully demonstrate the climbing of the soft robot on semi-smooth surfaces, including both sandpaper (Fig. 6a(iii) and Supporting Video S3) and indoor decoration painting wall (Supporting Video S3) carrying a load of 200g. It should be noted that to promote their contact and adhesion to the semi-smooth climbing surfaces, we did some treatment to flatten the bottom surface of the adhesion actuators (Fig. S6). This flatten-treatment is necessary and required, especially for climbing on semi-smooth substrates since it can increase the direct contact[40, 41] of the bottoms of adhesion actuators on a target surface. Thus, it can prevent air leaking and increase the conformability of the adhesion actuator on foreign surfaces, as well as increase the friction between the soft machine and the substrates (Fig. S6, see more discussions in the Supporting Information).

*Climbing and walking under water*

Last, we demonstrate the potential application of the soft adhesion actuator to design underwater walking and climbing soft robots. The maximum shear adhesion force of the adhesion actuator on a glass surface under water is measured to be ~10.62N (following the characterization method shown in Fig. S7) when subjected to 62kPa pressurization, which is even larger than its dry adhesion of ~7.51N on glass surface in air. The reason for the larger adhesion under water may be due to the fact that the increase in the cavity volume will pull the water inside in tension, resulting in a decrease in the internal water pressure[27, 42]. This pressure drop in liquid may generate a larger pressure difference between the ambient and the cavity when compared to the actuator working on ground, thus leading to a firmer attachment of the adhesion actuator. The exact working mechanism for the enhanced adhesion under water will be studied in more details in the future. With this improved adhesion, we successfully demonstrate that the soft robot can walk (Fig. 7a, Supporting Video S4) and climb (Fig. 7b, Supporting Video S4) smoothly under water (limited to smooth surfaces) on the glass surface with a certain amount of loading capability.

**Conclusion**

In conclusion, we designed a bioinspired simple, novel proof-of-concept amphibious soft robot that can walk horizontally and climb vertically on different types of smooth, semi-smooth, dry, wet, or slippery substrates with a certain load-carrying capability on ground and under water. The success lies in our new design of soft adhesion actuators with embedded spiral pneumatic channels for switchable, strong, and mechanically robust (Fig. S8) adhesion on different types of surfaces upon pressurization. The soft adhesion actuator provides a new platform for designing soft robots that can operate on vertical surfaces and work under water, which could find potential applications in design of switchable adhesion materials, object transportation[11], wall-cleaning[43], camouflage machine[12], and underwater soft machines etc.

Despite the promise, some limitations of this work exist. One limitation of the current ACSR is that it only exhibits single degree-of-freedom in motion by using the classic pneumatic bending actuator[1, 38]. Since the aim of this research is to realize the climbing of a robust soft robot on vertical surfaces, we focus on the adhesion behavior of the proposed soft adhesion actuator rather than developing a high degree-of-freedom driving actuator. However, the driving actuator is crucial for soft robots because it determines the gaits and locomoting efficiency of the soft robot. Therefore, ongoing work will aim at modification of ACSR to realize faster speed and more locomotion types (e.g. turning, locomoting on ceiling and switching between different dimensions etc.) using a driving actuator with 3D mobility.

**Supporting Information**

Supporting Information is available online from Soft Robotics

**Acknowledgements**

J. Y. acknowledges the funding support from the start-up at Temple University. The authors thank the help from Yao Zhao, Bosen Qian, Connor O'Rourke, and Stephen H. Doroba in the set-up of locomotion testing.

**Author Disclosure Statement**

No competing financial interests exist.

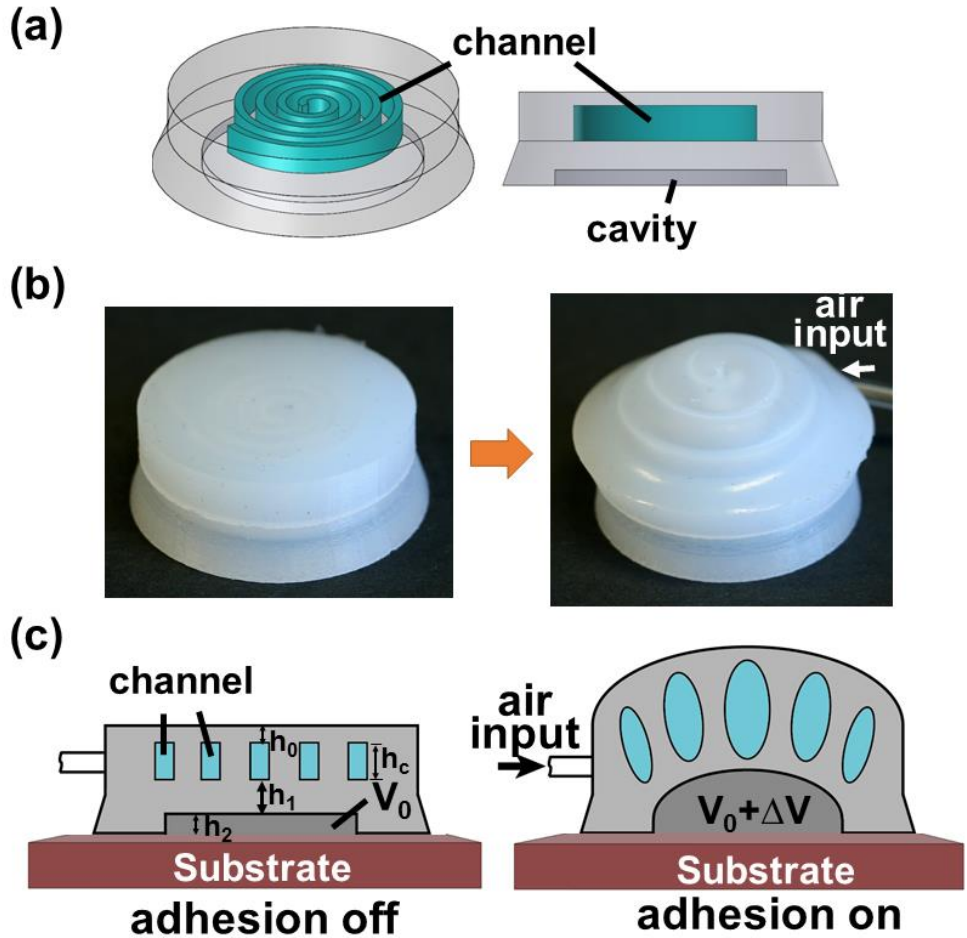

**Fig. 1.** Design of soft adhesion actuators. (a) Schematic design of the bilayer adhesion actuator with embedded spiral-shape pneumatic channel on the top and a cylindrical chamber or cavity underneath. (b) The as-fabricated adhesion actuator (left) deforms into a dome-shape upon pressurization in the air channel (right). (c) Schematic of the mechanism for switchable adhesion in the adhesion actuator upon pneumatic pressurization. Left: adhesion-off state, Right: adhesion-on state.

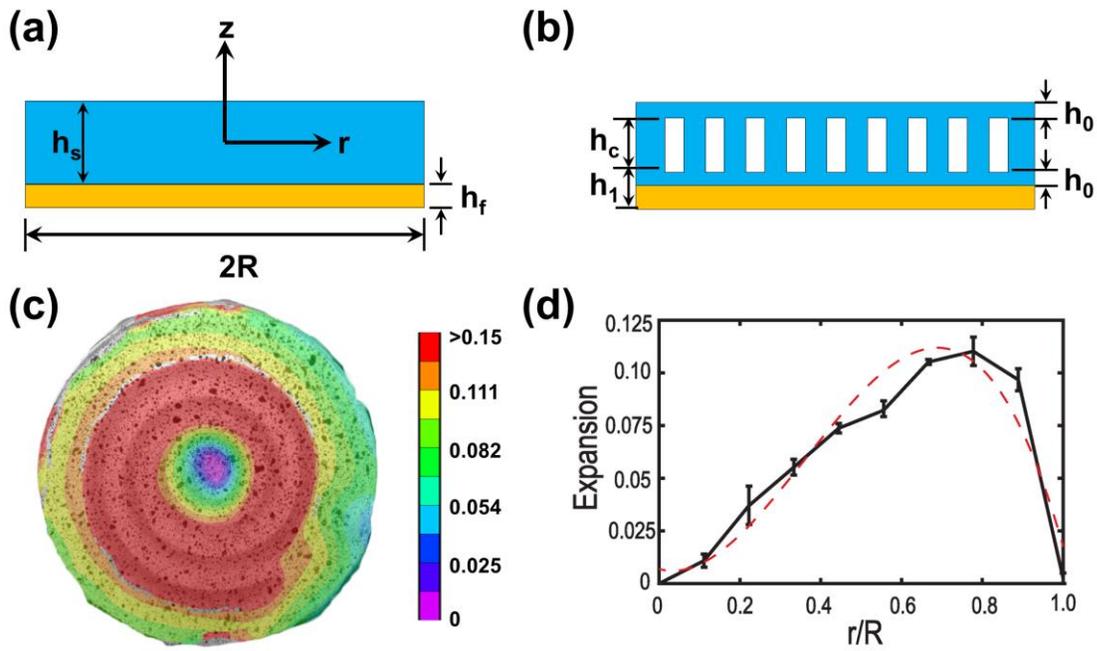

**Fig. 2.** (a) Schematic of continuum thin film/substrate bilayer system with misfit strain for deforming into a doming shape. (b) Schematic of the proposed bilayer doming model with pneumatic spiral channel. (c) DIC test of the channeled layer (indicated by blue in (b)) upon 4mL inflation shows the radial strain contour. (d) The measured expansion rate of the top layer along the radial direction.

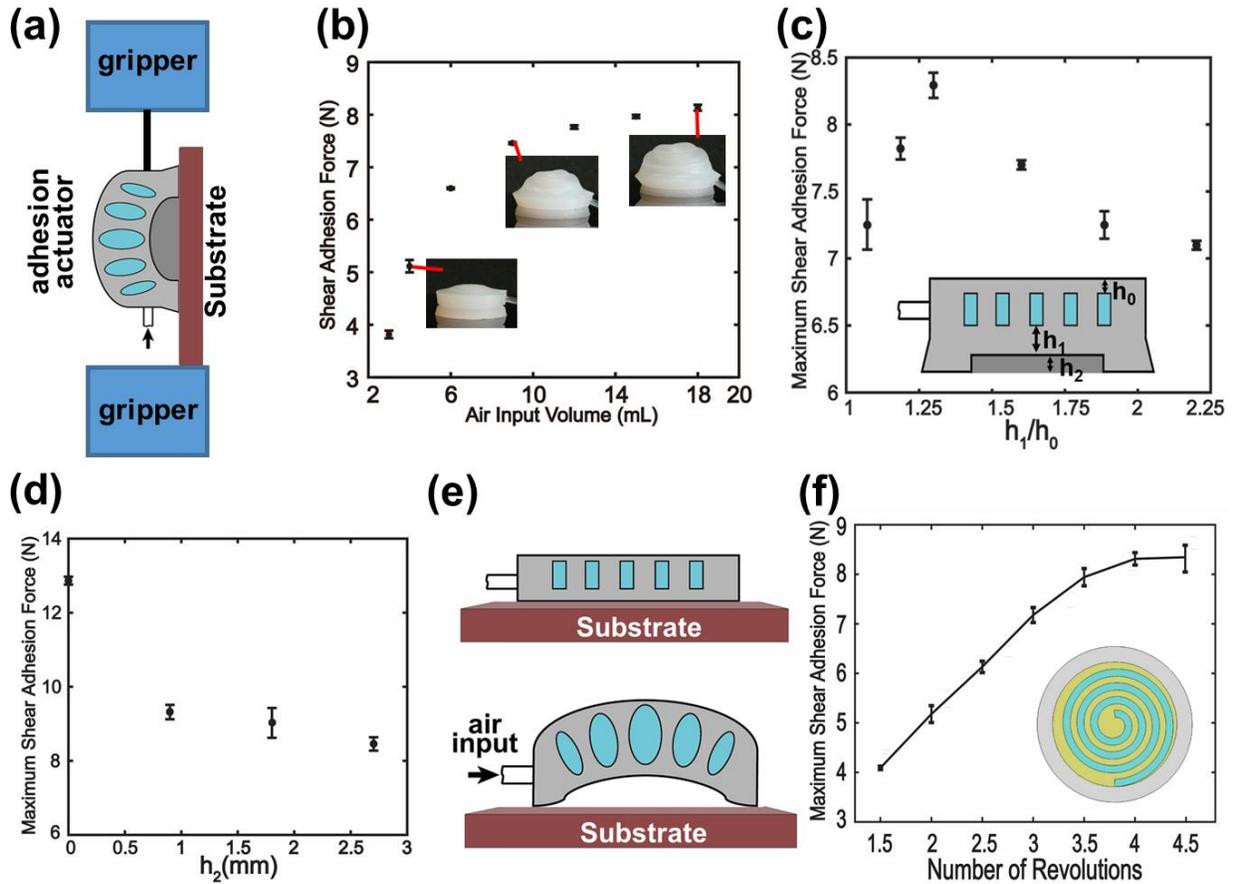

**Fig. 3.** (a) Schematic of shear adhesion test of the adhesion actuator. (b) The measured shear adhesion force as a function of the input air volume. (c-d) The measured maximum shear adhesion force (pressurized at 62kPa) attached to acrylic surfaces by varying $h_1/h_0$ ($h_2 = 2.7$mm) and $h_2$ ($h_1/h_0 = 1.35$) as illustrated in Fig. 1c. (e) Schematic of the adhesion actuator without cavity underneath the channel (i.e. $h_2=0$), showing less contact with the substrate upon pressurization. (f) The maximum shear adhesion force (pressurized at 62kPa) vs. the number of revolutions of the spiral structure in a unit volume (yellow).

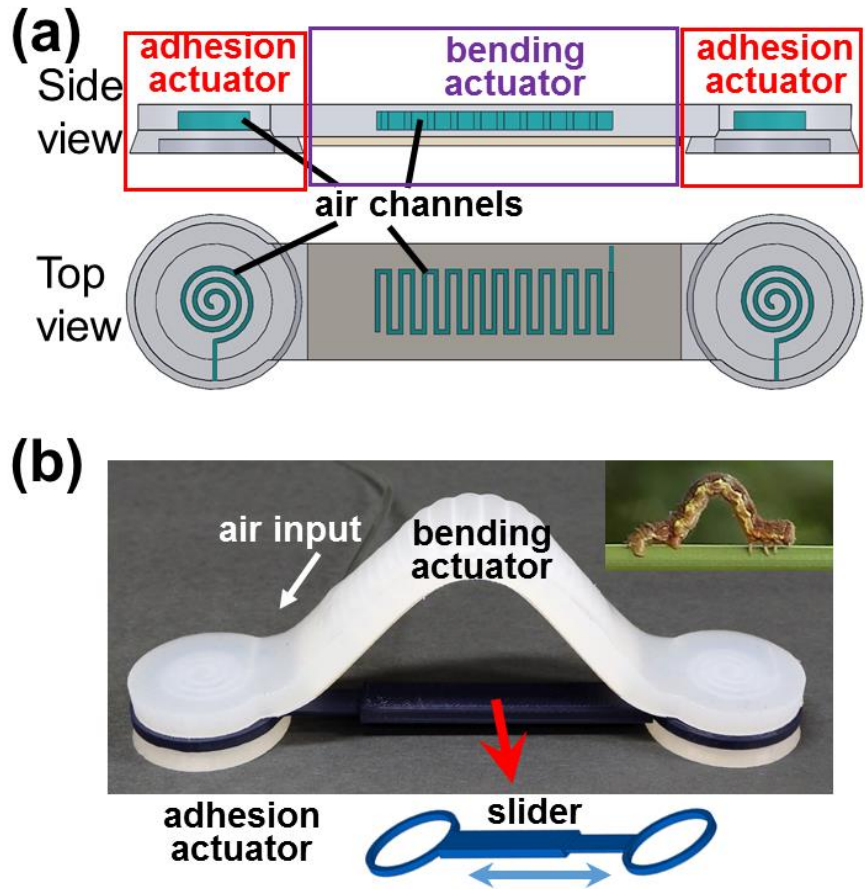

**Fig. 4.** Design of an amphibious climbing soft robot (ACSR). (a) Side view (top) and top view (bottom) of the schematic design of the ACSR composed of two adhesion actuators (two sides with embedded spiral channels) for switchable adhesion, and one bending actuator (middle with rectangular wave-like channels) for locomotion driven by pneumatic pressurization. (b) A fabricated ACSR from the design in (a) under a bended state upon pneumatic pressurizing the bending actuator to mimic the locomotion of an inchworm (top right inset).

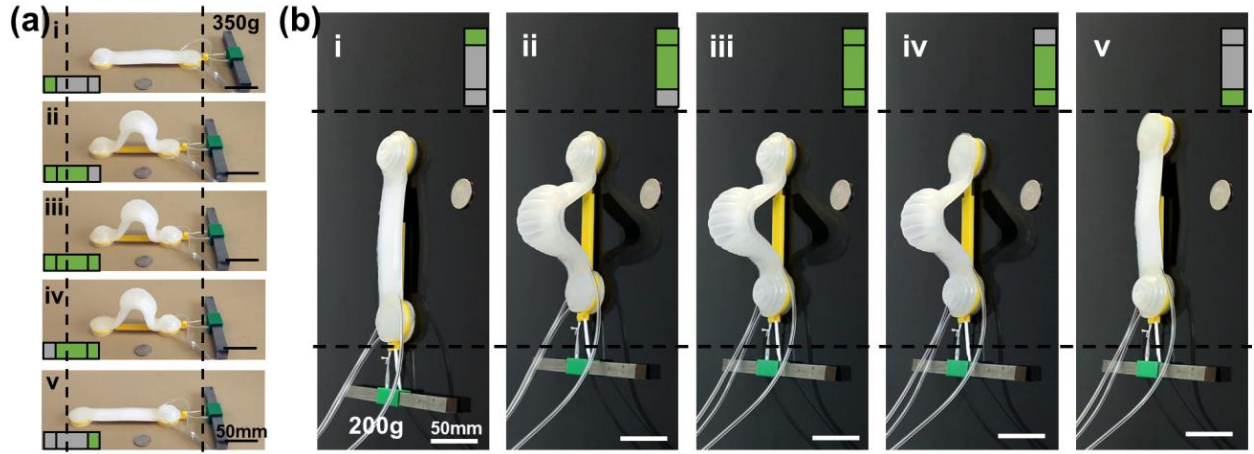

**Fig. 5.** Demonstration of the walking and climbing mode of ACSR with a carried load on ground. (a) snapshots of its walking mode on a smooth acrylic surface with a carried load of a 350g weight by actuating the three adhesion and bending actuators in 5 sequential steps from (i) to (v). In the bottom left inset, green color indicates the pressurization into the channel while grey color represents the depressurization, correspondingly, either the adhesion or bending actuator is on and off upon actuation or de-actuation. (b) Demonstration of the sequential actuating adhesion and bending actuators for vertical climbing on an acrylics surface with a carried load of 200g.

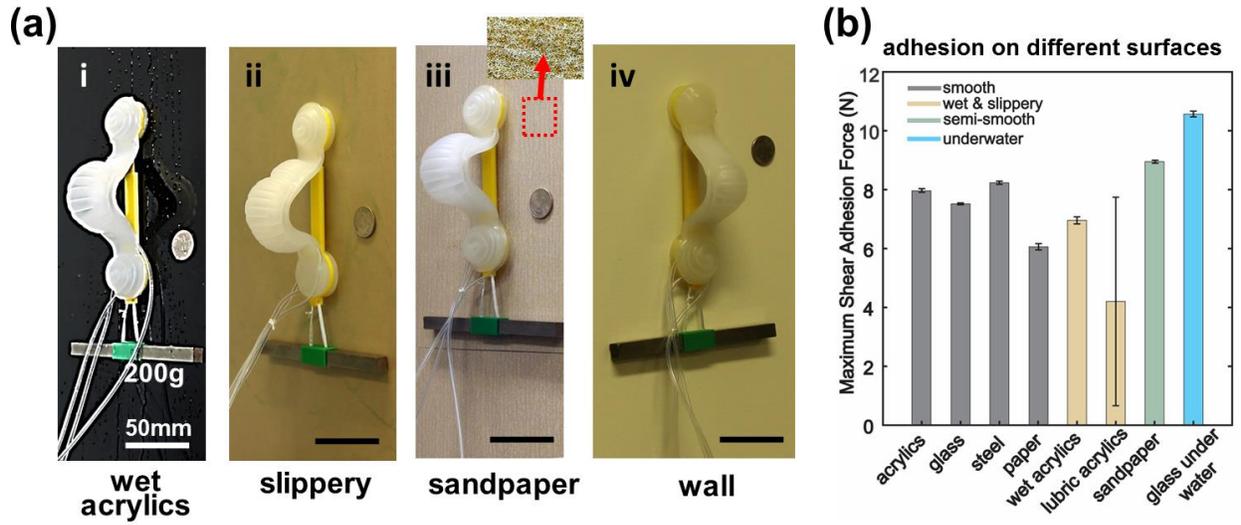

**Fig. 6.** Characterization of adhesion force and demonstration of vertical climbing of ASCR on different substrates with carried load. (a) Demonstration of the soft robot's wide capability of climbing on different types of vertical surfaces with a carried 200g load, (i) wet acrylics, (ii) lubricated slippery acrylics, (iii) semi-rough sandpaper, (iv) interior painted wall. (b) Summarization of the measured maximum shear adhesion force upon the same pressurization on different types of substrates.

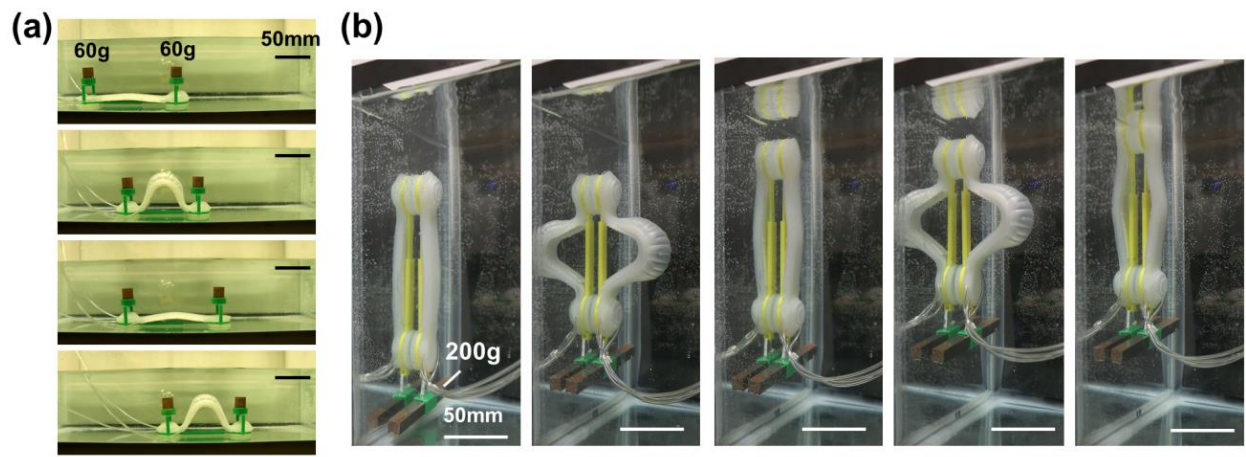

**Fig. 7.** Demonstration of ACSR's capability for underwater climbing and walking on glass. (a) Walking under water on a smooth glass; (b) Climbing a vertical glass wall under water with a carried load of 200g.